\documentclass{article}
\usepackage{graphicx} % Required for inserting images
\usepackage{color}
\usepackage{authblk}

\title{Network and Kinetics-based Biosignatures: Implications for the Putative Habitable World Observatory Design}
\author[1]{Theresa Fisher}
\author[2]{Chester Harman}
\author[3]{Estelle Janin}
\author[4]{Megan Shabram}
\author[5]{Shang-Min Tsai}
\author[6]{Nicholas Wogan}
\author[7]{Michael L. Wong}
\affil[1]{Steward Observatory, University of Arizona}
\affil[2]{Planetary Systems Branch, NASA Ames Research Center}
\affil[3]{School of Earth and Space Exploration, Arizona State University}
\affil[4]{Giant Magellan Telescope}
\affil[5]{Department of Earth and Planetary Sciences, University of California Riverside}
\affil[6]{Space Science Division, NASA Ames Research Center}
\affil[7]{Earth and Planets Laboratory, Carnegie Institution for Science}
\affil[8]{Corresponding author: theresafisher@arizona.edu}

\begin{document}

\maketitle

\section{Abstract}

The proposed Habitable Worlds Observatory is intended to observe the atmospheres of nearby terrestrial exoplanets with a resolution greater than that of any previous instrument. While these observations present a substantial opportunity for astrobiology, they also incur the risk of false positives and false negatives. Here, we explore the use of systems science (in the form of network theory and thermochemical kinetics) to mitigate these risks, and briefly describe the technical specifications HWO would require in order to use these methodologies.

\section{Introduction} \label{introduction}
%evaluate the role of complex interactions in the atmospheric chemistry of terrestrial exoplanets as a potential biosignature for the Habitable Worlds Observatory, and determine what minimum technical requirements might be required for the HWO to make use of these biosignatures.
%biosignatures grounded within complex system science approaches - more holistic than one-off biosignatures

The ongoing search for life beyond Earth hinges on our understanding of life and its origins on Earth. Historically, this has resulted in the identification of individual features that are driven wholly or in large part by terrestrial biology as candidate biosignatures, including the presence of atmospheric gases like oxygen (O$_{2}$), ozone (O$_{3}$), and methane (CH$_{4}$) or surface reflectance features like the vegetative red edge of chlorophyll absorption \cite{schwieterman2018exoplanet}. However, these features do not occur in a vacuum nor arise spontaneously, which in turn demands contextual measurements to underpin the environmental setting and rule out potential abiotic mechanisms that could produce the observed features. To first order, for example, the simultaneous observation of O$_{2}$ and CH$_{4}$ is taken as a robust biosignature pair because the coincidence of these two species requires large source fluxes to maintain both species, which would otherwise react and drive one or both to near-zero concentrations. 

Any scientific claim, whether it be for the presence of volcanoes \cite{kaltenegger2010detecting,misra2015transient}, continents \cite{fujii2017rotational,cowan2017mapping}, or a habitable or inhabited planetary surface \cite{schwieterman2018exoplanet}, will only be as strong as the combination of evidence and explanation from established theories. This biases us towards biosignatures with a well-understood mechanism that behaves predictably under a broad range of conditions. Given the current state of astrobiology theories, the logical extension of this rationale is that no biosignature can exist in isolation \cite{lovelock1965physical}, and, in fact, the most robust biosignature interpretations are supported by the suite of all possible contextual, related, and parallel observables \cite{catlingdavid2018exoplanet}.

The reality of the situation, however, is that exoplanet observations individually only convey a small slice of truth, and each additional parameter or measurement carries with it its own uncertainty. Scientific inference incorporates uncertainties beyond the raw data obtained by the instrumentation. A retrieved abundance depends on the chemical network and planetary environment model, e.g., as part of an atmospheric spectral retrieval scheme used in an analysis pipeline, highlighting that uncertainties and unknowns in chemical and physical properties and processes will affect the final interpretation. The compounding uncertainty associated with all these elements has parallels with the target observables themselves, meaning that a systems science approach can provide well-supported and exhaustive insights into not only each individual element but how likely it is to observe them together. 

Below, we detail case studies that highlight several existing methods that leverage systems science to strengthen biosignature detections. Observations of the composition of planetary atmospheres can be examined to look for disequilibrium between constituents, and diagnose tendencies in the chemical networks, or to identify anomalously complex species. All of these approaches rely on how well we understand fundamental physicochemical and spectroscopic properties, and can be used to identify gaps in our knowledge that should be addressed as part of our preparation for the search for life elsewhere. Ultimately, no observation or instrument stands alone, and only with thorough preparation and a holistic systems approach will we find confidence in our interpretations.

%What networks and kinetics research is required to improve the robustness of the evidence that HWO gathers for the presence of life elsewhere and the explanations required to support those measurements?

\section{Case Studies} \label{case_studies}

\subsection{Atmospheric Photochemical Disequilibrium} \label{atmospheric_disequib}

A more holistic approach to atmospheric biosignatures is examining the thermodynamics of the entire atmosphere, first suggested by Lovelock in 1965 \cite{lovelock1965physical}, in terms of looking for atmospheric chemical disequilibrium. Measures of disequilibrium have subsequently been further refined, including the power required to sustain the disequilibrium \cite{simoncini2013quantifying}, and the total available Gibbs energy, $\Phi$, in the atmosphere \cite{krissansen2016detecting}. 

\begin{figure}
    \centering
    \includegraphics[width=1\linewidth]{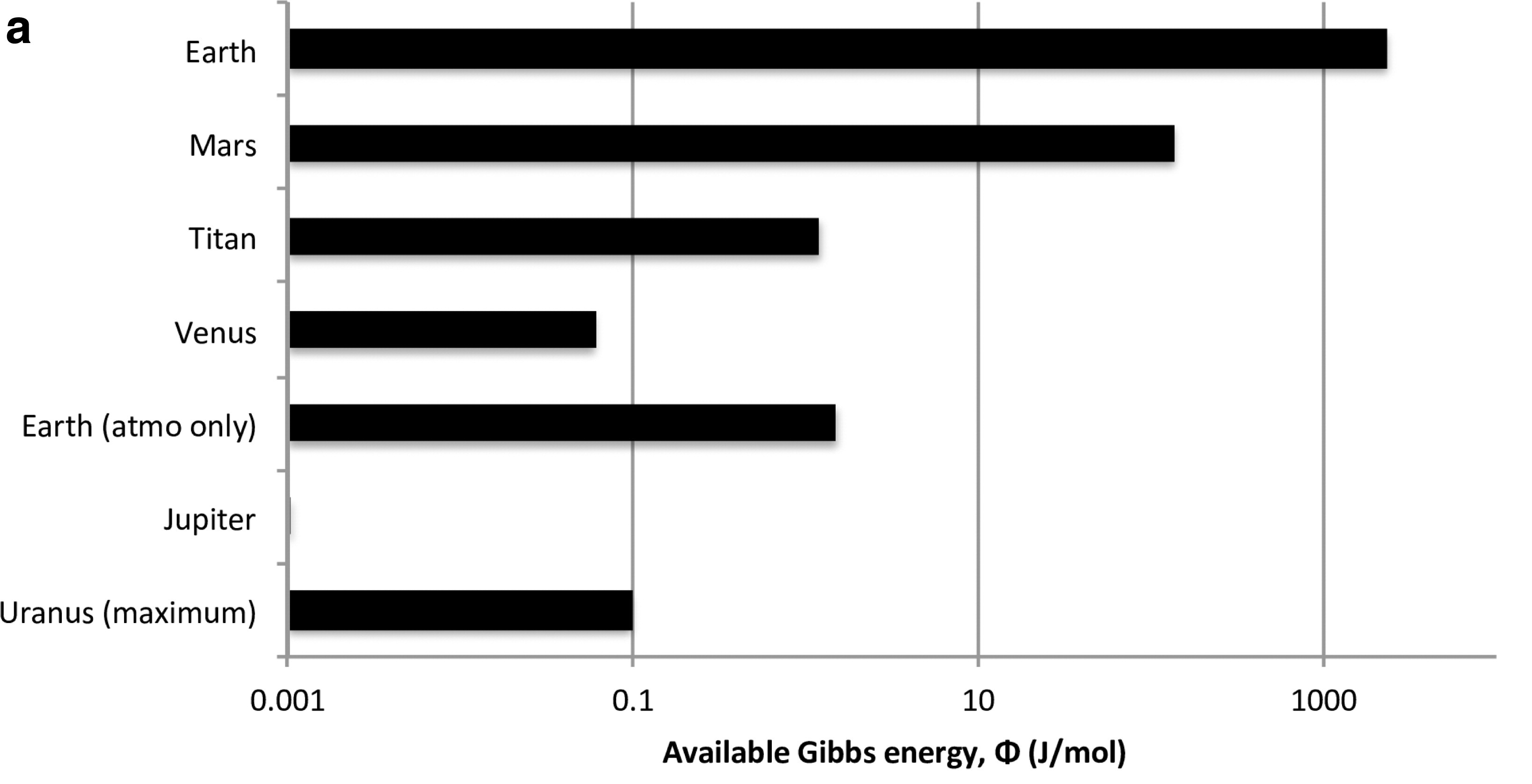}
    \caption{Comparison of the available Gibbs energy, $\Phi$ in Solar System atmospheres from Krissansen-Totton \textit{et al.} 2016 \cite{krissansen2016detecting}}.
    \label{fig:phi-measurements}
\end{figure}

\begin{figure}
    \centering
    \includegraphics[width=1\linewidth]{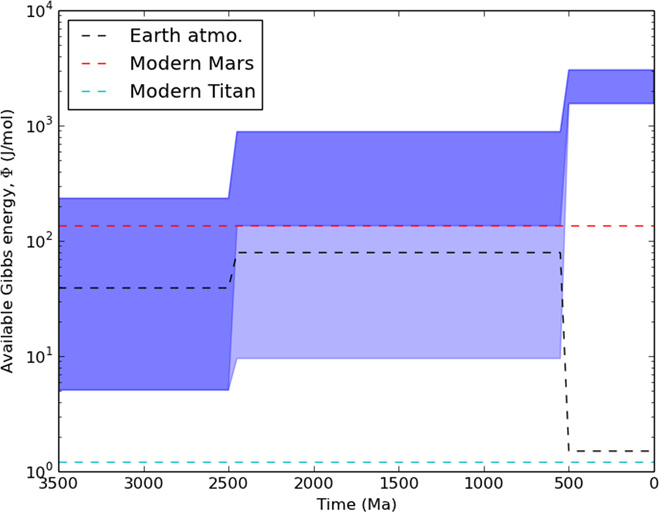}
    \caption{The evolution of Earth’s atmosphere-ocean disequilibrium through time, as measured by available Gibbs energy, from Krissansen-Totton \textit{et al.} 2018 \cite{krissansen2018disequilibrium}. The blue shaded regions show the evolution of Earth’s atmosphere-ocean disequilibrium, with the wide range being due to the large uncertainties in the amount of O$_2$ present. The black dashed line represents the Earth's maximum available Gibbs energy when measured single-phase, i.e. only accounting for the chemistry of the atmosphere and not the ocean. The available Gibbs energy of Mars' and Titan's atmosphere are shown for comparison.}
    \label{fig:phi-multiphase}
\end{figure}

The latter was later modified to account for multiphase equilibrium \cite{krissansen2018disequilibrium} by integrating ocean chemistry. The available Gibbs energy is determined by calculating the atmospheric and oceanic composition at chemical equilibrium and then comparing it to the actual observed values. When applied to planetary atmospheres within our solar system, Earth's atmosphere is easily distinguishable, having an order of magnitude or more available Gibbs energy than the next closest planet \cite{krissansen2016detecting} (see Figure \ref{fig:phi-measurements}), and this number may have been even higher in Earth's past in the run-up to the Great Oxidation Event \cite{krissansen2018disequilibrium} (see Figure \ref{fig:phi-multiphase}).

While these techniques require being able to infer the presence or absence of an ocean on a planet's surface to be of full efficacy, modeling of HWO's capabilities suggests that ocean glints may be detectable in as many as 10\% of star systems observed by the telescope \cite{vaughan2023chasing}.

\subsection{Atmospheric Chemical Reaction Network Topology} \label{atmospheric_CRNs}

\begin{figure}
    \centering
    \includegraphics[width=1\linewidth]{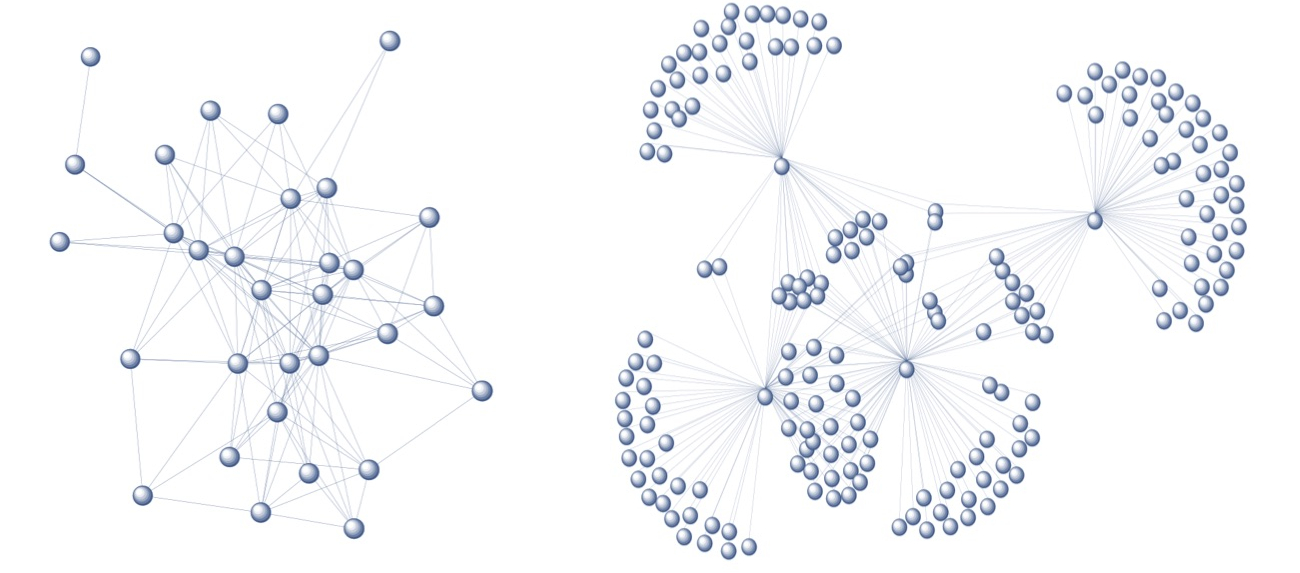}
    \caption{Simplified examples of atmospheric chemical reaction networks. The network on the left represents the topology typically seen in most planetary atmospheres, such as Mars; on the right, a network modeled after Earth's atmospheric chemical reaction network.}
    \label{fig:example_CRNs}
\end{figure}

Another approach to analyzing the atmospheric chemistry of a planet is to represent it as a network, with the species represented as points, or nodes, which are then connected via lines, or edges, based on what reactions they co-participate; these edges can then be weighted by the reaction rate of the reactions or other known chemical parameter (see Figure \ref{fig:example_CRNs}). The topological properties of this network can then be measured and analyzed \cite{fisher_inferring_2022} to aid in understanding the behavior of the atmospheric chemistry. For example, mean degree ($k$), which refers to the number of connections or edges attached on average to a given node, can be interpreted as the average of the total flux driven by each chemical compound across the entire network. 

Initial research suggests that the topology of Earth's atmosphere is unique in the solar system, possessing a lower mean degree and higher average shortest path length than any other planetary body \cite{sole2004large,wong2023toward}. Additionally, this topology resembles that of biochemical networks more than astrophysical ones \cite{jolley2012topological,wong2023toward}. Combined, these results suggest that the presence of a biosphere can influence the characteristics of a planet's atmospheric chemical reaction network, and that this influence can serve as a potential biosignature, and in some cases, provide a higher confidence detection than just atmospheric abundances alone \cite{fisher_complex_2023}. 

\begin{figure}
    \centering
    \includegraphics[width=1\linewidth]{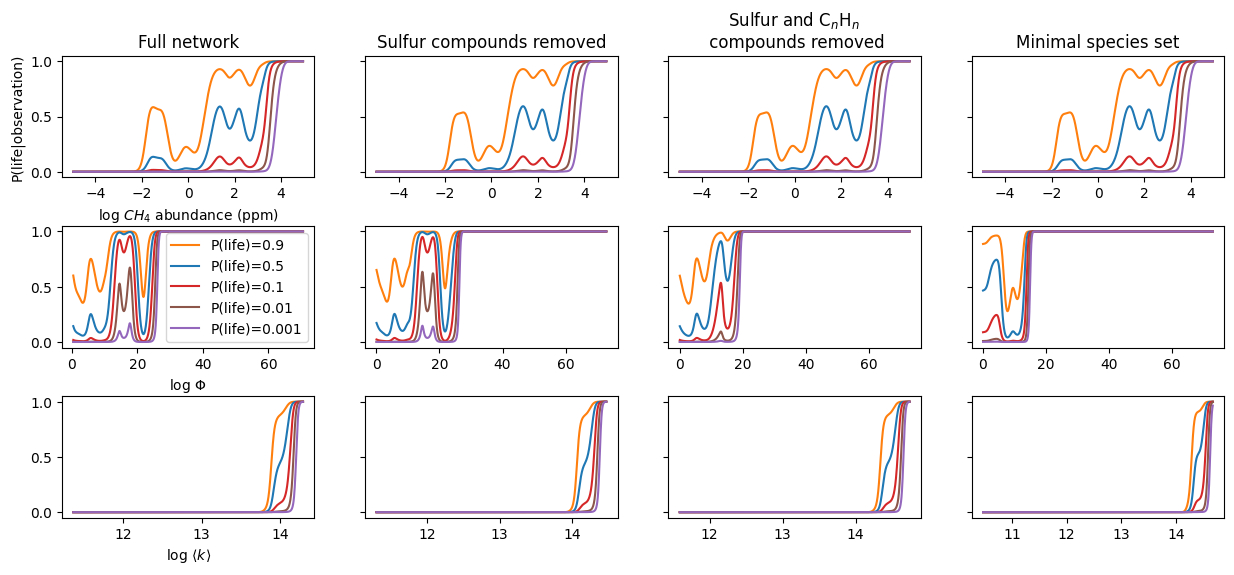}
    \caption{The probability of life detection given a set of observations, where \textit{P(life)} is the probability of life on a given planet. Observations were drawn from a set of 10,000 simulated atmospheric models, half incorporating a methanogenic biosphere, the other half purely abiotic, and used as input for a Bayesian analysis as \textit{P(observation)}. From left to right: Probabilities of detection calculated from the full atmospheric chemical reaction network; probabilities calculated with sulfur compounds removed from the atmospheric chemical reaction network; probabilities calculated with both sulfur and hydrocarbons (C$_n$H$_n$, excluding methane) removed; probabilities calculated using a minimal chemical reaction network composed of only H$_2$, CO, CO$_2$,CH$_4$, and O$_2$. In all cases, mean degree ($k$) and phi ($\Phi$) yield higher confidence over a wider range of \textit{P(life)} than the abundance of CH$_4$ alone. }
    \label{fig:methods comparison}
\end{figure}

As an example, analysis of a set of 10,000 early Earth 1-D atmospheric models produced for this paper using PyAtmos \cite{chopra2023pyatmos} suggests that network topology can provide high confidence biosignature detection even when a relatively small number of species have been identified (see Figure \ref{fig:methods comparison}). In this case, half of the models incorporated a methanogenic biosphere, while the other half only included abiotic production of methane. Once the models converged, the final abundances of their species was used to construct a reaction-rate-weighted chemical reaction network that was then analyzed; the available Gibbs free energy ($\Phi$) was also calculated (after \cite{krissansen2016detecting}; see Section \ref{atmospheric_disequib} for more detail). This output was then used to calculate $P(life|observation)$ (the probability of life being accurately detected) given \textit{P(observation)} (a set of observations) and a range of values for \textit{P(life)} (the prior probability that life exists elsewhere in the universe):

\begin{equation}
    P(life|observation) = \frac{P(observation|life)(P(life)}{P(observation)}
\end{equation}

Both $k$ and available Gibbs free energy ($\Phi$) yielded higher confidence over a wider range of $P(life)$ (see Figure \ref{fig:methods comparison}) than just the atmospheric abundance of CH$_4$ alone, even when the number of species in the modeled atmosphere was considerably reduced (see Section \ref{statistical_approach} for more information concerning the use of Bayesian probability in this context).

\subsection{Statistical Approach} \label{statistical_approach}

One way to approach to the inherent uncertainty in exoplanet observations is through statistical population analysis techniques. An example approach is Hierarchical Bayesian Modeling (HBM), which models latent variables in a probabilistic framework that accommodates uncertainty at multiple levels (e.g., individual-level variations and group-level variations). These two statistical approaches could be deployed in a complex systems framework to link observables with hidden clusters, classes, or groups of co-evolutionary trajectories. As one example, quantifiable chemical reaction network measurements (e.g., average shortest path length) can differentiate network types. 

When developing a framework for applying a probabilistic population model such as HBM analysis to exoplanet atmosphere data, considering instrumental limitations and uncertainties can inform model assumptions. \textit{In reverse, understanding model inputs and assumptions can also drive instrument design and development.} 

For example, randomized network comparisons show distinct differences from observed biological networks \cite{kim2019universal}. We suggest to focus on Earth-like  networks when analyzing exoplanet data, and start with a simple statistical model that explores categorizing exoplanet atmospheres based on network topology between two network topology classes, i.e., Earth-like vs. Mars-like, via a two-component Gaussian mixture model. This framework has the potential to accommodate the need to account for and propagate measurement uncertainties from both laboratory and instrumental sources, and track model uncertainty factors. Abstraction layers and increased model complexity can be carefully considered such as assumptions about correlations between evolutionary tracks and host stars, to further scientific yields through these statistical approaches \cite{shabram2016eccentricity}. Another example approach is the application of an HBM occurrence rate framework similar to what has been demonstrated in \cite{shabram2020sensitivity}, and could support the exploration of how rare or common Earth-twin network topologies are among exoplanets.

\section{Technical Plausibility and Requirements}
While these approaches are agnostic to the specific biochemistry of a putative alien biosphere, they rely on our understanding of chemistry, i.e., how each node representing a species in a chemical reaction network are interconnected, or the thermodynamics of the underlying reactions. This requires at least some degree of inference of atmospheric chemistry, either via photochemical modeling or direct inference from spectral data. As a result, certain technical requirements still remain for these approaches to be viable. Foundationally, we will need to be able to identify, at a minimum, at least 3 or more species in a planet's atmosphere--optimally H$_2$O, O$_2$, O$_3$, CH$_4$, and CO$_2$, as there exists a wealth of literature on placing these gases in context as biosignatures\cite{schwieterman2018exoplanet}--and the resulting chemistry that could evolve from interactions between those species given the thermodynamic environment. Below, we briefly summarize these requirements. 

\subsection{Kinetics requirements}
Given the important role photochemical kinetics play in the analysis methods, and the uncertainties associated with the photochemical reaction rates in both Earth-like \cite{thompson1991effect} and non-Earth-like \cite{hebrard2006photochemical} atmospheres, it would be worthwhile to measure selected rates in a laboratory environment, and so increase precision. 
\subsection{Spectral requirements}
The opacities in the UV and the infrared are both key for evaluating the habitability of an atmosphere. The UV cross sections are essential for photochemical modeling, whereas the infrared line lists dictate how we translate the spectral data into molecular features.

In particular, H$_2$O and CO$_2$ are two key molecules concerning UV cross sections. Measurements of H$_2$O's near-UV cross sections have only been conducted at room temperature, revealing some very weak photodissociation cross sections at 240 nm. The small cross section leads to the light penetrating deeper into the atmosphere so H$_2$O will dissociate much deeper and initiate chemistry therein. 

Additionally, a broad range of wavelengths can enable detection of potentially more species, which will aid in using these metrics, and in modeling the atmospheres in question more generally.

\bibliography{cited}

\end{document}